# The VLT dealing with the Atmosphere, a Night Operation point of view

Julio Navarrete

European Organization for Astronomical Research in the Southern Hemisphere
Alonso de Cordova 3107, Vitacura, Santiago-Chile

## ABSTRACT

The Science Operation Department is composed of Astronomers, Telescope Instruments Operators (TIO) and Data Handling Administrators (DHAs). Their main goal is to produce top-quality astronomical data by operating a suite of 9 telescopes, 14 Instruments and related systems, supporting the execution of Visitor Mode Observations or executing Service Mode Observations. Astronomers and TIOs have to deal with atmospheric parameters like seeing, coherence time, isoplanatic angle, precipitable water vapor, etc. in order to take in real time the best decisions on the best program to be executed according to the current conditions. We describe the tools available in the control room, provided by the environmental monitoring and forecast system.

**Keywords:** VLT, atmosphere, seeing, coherence time

## 1. INTRODUCTION

The Very Large Telescope array (VLT) is the flagship facility for European ground-based astronomy at the beginning of the third Millennium. It is the world's most advanced optical instrument, consisting of four Unit Telescopes with main mirrors of 8.2m diameter and four movable 1.8m diameter Auxiliary Telescopes. The telescopes can work together, in groups of two or three, to form a giant 'interferometer'[1]

The atmosphere plays a critical role in the operation of this great machine called VLT so the atmospheric parameters information (past and present) we can get using the tools developed during the site testing and available for the observatory are widely used for the observations which are carried out every night.

The Observatory offers two observing modes, Visitor Mode and Service Mode, the latter being where the Astronomers and TIOs should use the tools available to plan a line of observations and to take in real-time decisions about the best program to run depending on the current conditions.

## 2. WHO ARE WE AND WHAT DO WE DO[2]

The Paranal Science Operation Department is composed by 27 Staff Astronomers, 15 postdoctoral Fellows, 19 Telescope Instrument Operators (TIO), 5 Data Handling Administrators and 1 Department Assistant and our tasks are:

• Produce top-quality astronomical data by operating a suite of 9 telescope and 14 instruments and related systems

• Maintaining & improving performances of instruments

• Monitoring instrument health

• Providing calibration files associated to science files

• Supporting preparation of astronomical observations (mainly in visitor mode)

• Delivering data to the user

• Working with engineering in troubleshooting, commissioning, technical intervention activities

• Developing, improving operations tools and procedures

• Interfacing with Garching partners: USD, QC, Instrumentation

# 3. OBSERVING AT PARANAL

Visitor Mode: The Astronomer goes to Paranal to carry out its own programs. We, the Science Operation staff support the preparation of the astronomical observations an execute them. There are not atmospheric constraints to carry out the programs that the visiting astronomer wants to observe (good luck, we wish you good seeing and clear skies)

Service Mode: Science Operation staffs (Astronomers and TIOs) are the ones in charge of to carry out the programs taking care of to accomplish the constraints set by the investigator.

The Night Astronomer has to plan the night by selecting the programs that better match the conditions (atmosphere) for the night. She/he uses the Observing tool (Fig. 1) that contains all the programs for all the instruments, to select the programs according to the ranking (A, B, C) and the constraints they have.

Figure 1. Observing tool showing the list of observations that can be observed this night

Once the programs are selected, they are displayed on the corresponding Instrument's queue with the constraints its have (Fig. 2).

Figure 2. Instrument queue showing the programs with the constraints (this panel is for SINFONI class A programs)

And finally they go into the execution sequence (Fig. 3)

Figure 3. Execution sequence panel showing the list of programs to be execute

Due to the large number of instruments (Fig. 4) covering wavelengths from ultraviolet to the mid infrared with and without adaptive optics facilities, the Night Astronomers and TIOs have to know the atmosphere behavior at least statistically in order to optimize the programs to be observed during the night.

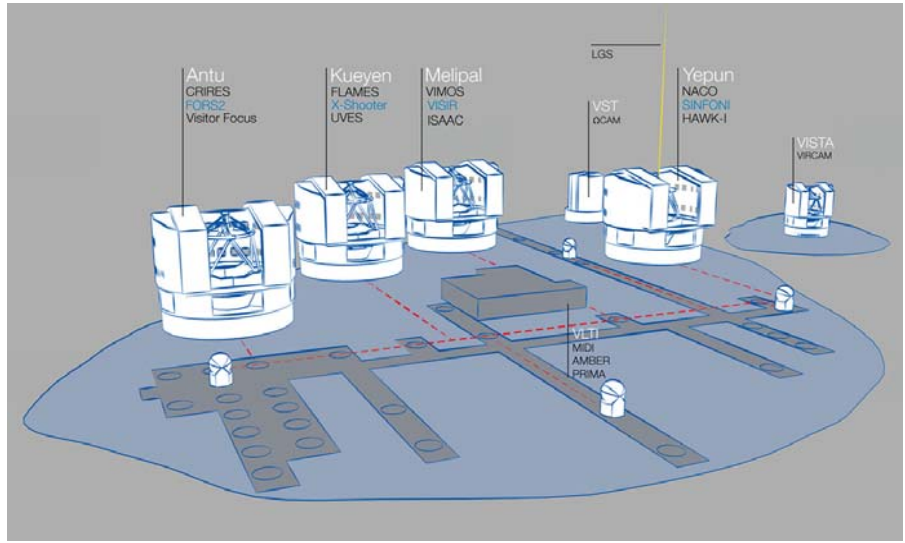

Figure 4. VLT and VLTI Array layouts showing the Instruments per Telescope.

**3.1 Instruments available in Paranal**

Due to the large number of instruments available in Paranal (Fig. 5), a brief description of them is given just as an information:

VLT Instruments[3]:

FORS (FOcal Reducer and Spectrograph) , is multi-mode instruments that can be used for imaging in the visible and for low-resolution spectroscopy.

ISAAC (Infrared Spectrometer And Array Camera) is a cryogenic infrared imager and spectrometer, observing in the 1 to 5 µm range.

UVES (Ultra-violet and Visible Echelle Spectrograph) is the high-dispersion spectrograph of the VLT, observing from 300 nm to 1100 nm, with a maximum spectral resolution of 110 000.

NACO is an Adaptive Optics facility producing images as sharp as if taken in space.

VIMOS (VIsible Multi-Object Spectrograph), a four-channel multiobject spectrograph and imager, allows obtaining low-resolution spectra of up to 1000 galaxies at a time.

FLAMES (Fibre Large Array Multi-Element Spectrograph) offers the unique capability to study simultaneously and at high spectral resolution hundreds of individual stars in nearby galaxies.

VISIR (VLT Imager and Spectrometer for the mid-InfraRed) provides diffraction-limited imaging at high sensitivity in the two mid infrared (MIR) atmospheric windows (8 to 13 µm and 16.5 to 24.5 µm).

SINFONI is a near-infrared (1 - 2.5 µm) integral field spectrograph fed by an adaptive optics module.

CRIRES (CRyogenic high-resolution InfraRed Echelle Spectrograph) provides a resolving power of up to 100 000 in the spectral range from 1 to 5 µm.

HAWK-I (High Acuity Wide field K-band Imager) is a near-infrared imager with a relatively large field of view.

X-SHOOTER (a wide-band [UV to near infrared] spectrograph) is designed to explore the properties of rare, unusual or unidentified sources.

VLTI Instruments:

MIDI is a MID-infrared Interferometric instrument for photometry and spectroscopy

AMBER is a near infrared Astronomical Multi-Beam combineR — an instrument for photometric and spectroscopic studies, which combines the light of three telescopes.

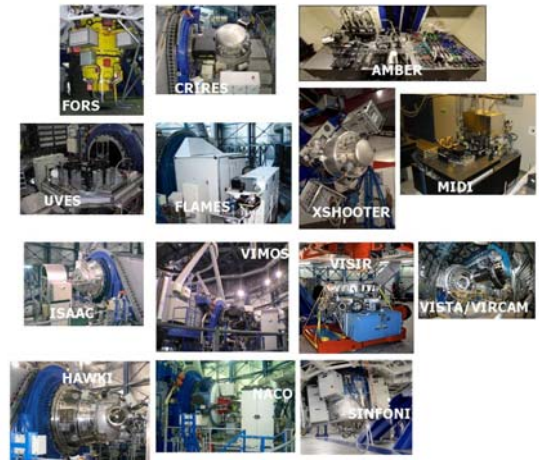

Figure 5. Instruments available at Paranal.

## 4. TOOLS AVAILABLE FOR ENVIRONMENTAL MONITORING AND FORECAST

To help Astronomers and TIOs to take the best in real time decisions, we count with a set of tools that permit us to monitor the atmosphere and to adapt the observations according to the actual conditions and with a set of forecasted parameters that allow us to get an idea about the night conditions in advanced

The Astronomical Site Monitoring (ASM) delivers in real time the local meteorological information like temperature, humidity and wind, etc., atmospheric information like seeing, coherence time, etc. The ASM, includes the all sky camera (MASCOT) for clouds monitoring. Figure 6 shows the display of all this information.

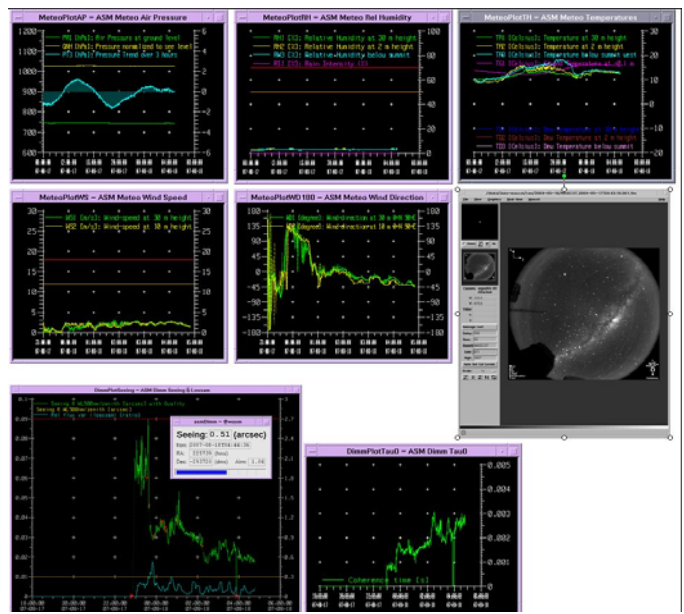

Figure 6. Astronomical Site Monitoring Display

The Forecast system is a web based tool[3] (Fig. 7) that give us short (36 hours with a step of 6 hours) and long term (8 days with a step of 12 hours) meteorological predictions at different pressure levels (Fig. 8).

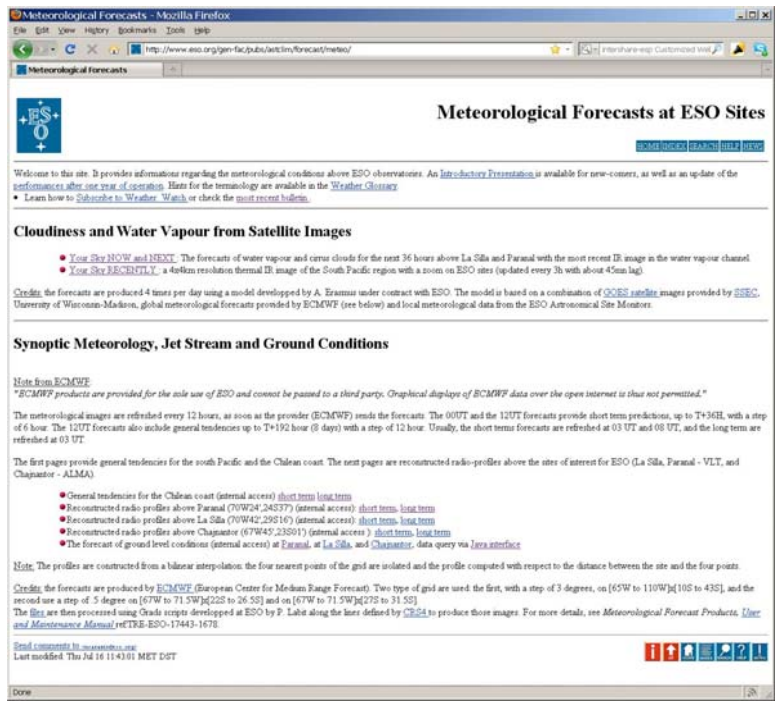

Figure 7. Web page for the Meteorological Forecast at ESO Sites

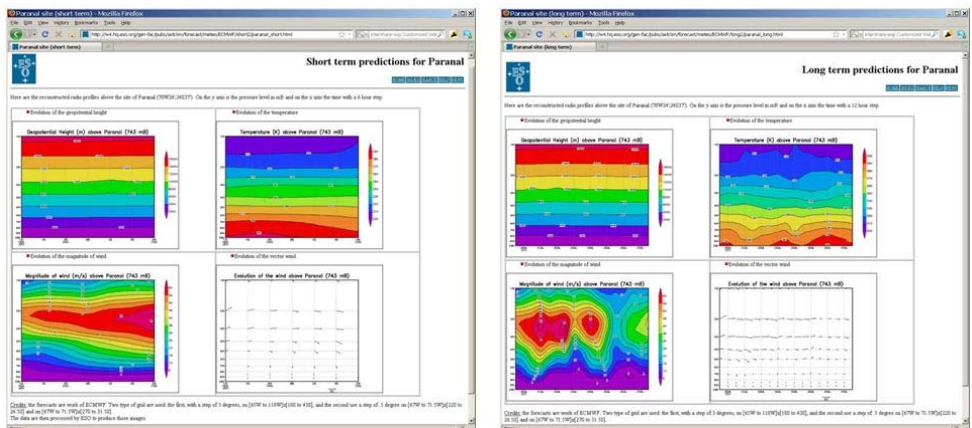

Figure 8. Short and long term prediction of the Geopotential height, Temperature and Wind above Paranal, y axis is the pressure level, x axis is the time (in hours for the short term and days for the long term) and the color bar is the value of the parameter.

Cloudiness and water vapor prediction for the next 36 hours (Fig. 9).

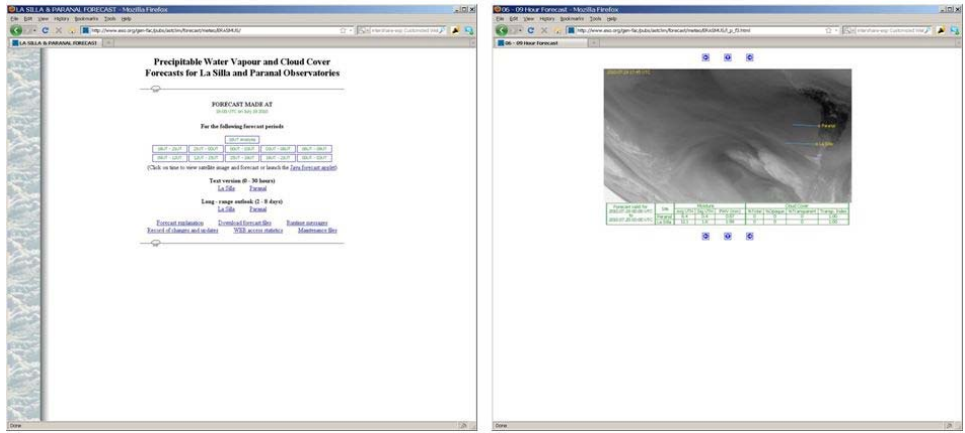

Figure 9. Precipitable Water Vapor and Cloud Cover Forecasts for ESO Observatories, at the right side of the figure, the page to select the forecast period to see and at the left side the results for the selected period.

All this information that can be see independently is compiling in a bulletin that is update 5 time per day for cloudiness and water vapor and 2 time per day for pressure, temperature and wind (Fig. 10).

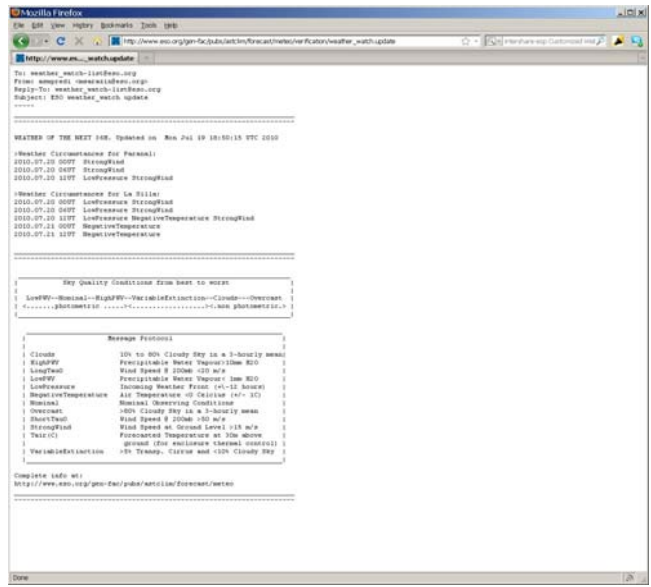

Figure 10. The bulletin with the next 36 hours weather circumstances

All these tools plus the knowledge of some characteristics behavior of some key parameters like seeing make life a little bit easy for Astronomers and TIOs. Paranal has a large atmospheric database that comes from the early VLT site testing, the analysis of parameters like seeing and coherence time with respect to some meteorological parameters like wind, allow us to identify some correlations that are useful when planning the observational queues for the night.

## 5. DEALING WITH THE ATMOSPHERE

The seeing is one of the most important parameters for all kind of observations (imaging, spectrography, interferometry, etc) with and without adaptive optics facilities, coherence time and isoplanatic angle are critical for observations with adaptive optics instruments, water vapor is very important for mid infrared observations. From the statistics of our database, we can set the following advices to manage the queue during the night.

## 5.1 Case good and bad seeing based on 5 years statistics (2005-2009)

The highest probability to get super seeing (seeing < 0.5") is when the wind comes from NNW-N with a speed of 2 to 8 m/s (Fig. 11), in this case, the Astronomer put programs that require this condition into the execution sequence. Usually, these are the hardest programs to execute and the highest ranked too, so in this case they have the highest priority to be executed

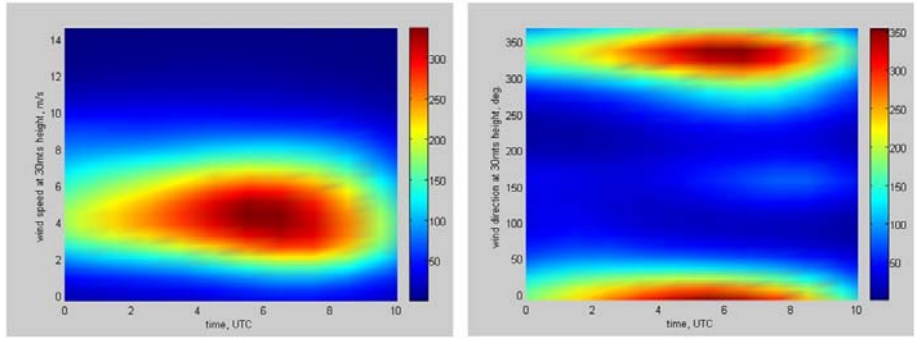

Figure 11. 2-D histograms showing the dependency of the good seeing (< 0.5") with the local wind at 30mts Height. Y axis is the wind speed in m/s for the left figure and wind direction in degrees for the right figure, X axis is the time in hours for the both figures and the color bar is the frequency of the good seeing.

Wind coming from NNE-NE (Fig. 12), usually brings bad seeing (seeing > 1.5"). Astronomers put programs with more relaxed constraints (class B or C programs) when they find this kind of conditions.

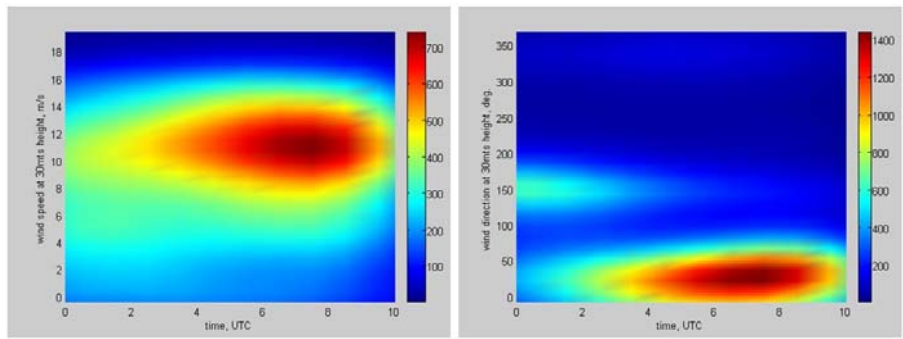

Figure 12. 2-D histograms showing the dependency of the bad seeing (> 1.5") with the local wind at 30mts Height. Y axis is the wind speed in m/s for the left figure and wind direction in degrees for the right figure, X axis is the time in hours for the both figures and the color bar is the frequency of the good seeing.

## 5.2 Case coherence time and isoplanatic angle based on 5 years statistics (2005-2009)

Astronomers and TIOs driving instruments with AO facilities know or should know about coherence time and isoplanatic angle too, as these are critical parameters to execute programs that require adaptive optics. We know that coherence time should be > 2.5msec to get the best performance of the adaptive optics instruments and larger the isoplanatic angle, larger the field to search for reference stars to correct the atmosphere. From the statistics, we know that the both parameters are highly correlated with the wind speed at 200mB (jet stream) and that the jet stream is very active during winter time (Fig. 13).

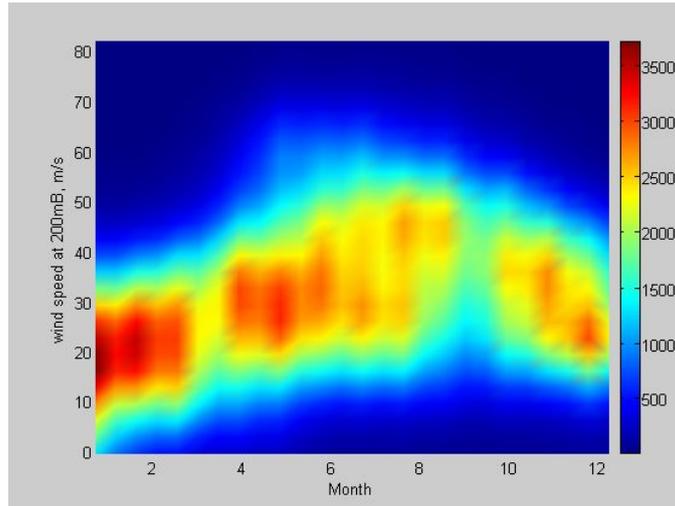

Figure 13. 2-D Histogram of the wind speed at 200mB as function of the months, Y axis is the wind speed at 200mB in m/s, X axis is month and the color bar is the frequency of the wind speed at 200mB.

Figures 14 and 15 show the correlation between coherence time and isoplanatic angle with the jet stream.

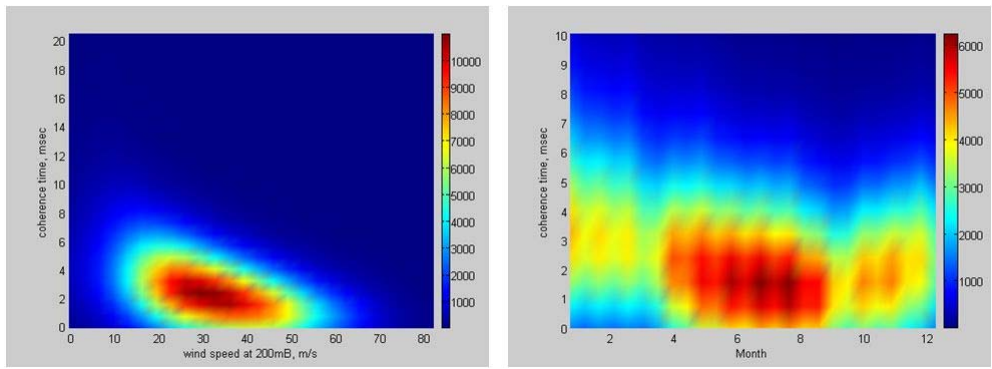

Figure 14. 2-D Histogram of the coherence time in msec as function of the wind speed at 200mB (left side) and the coherence time as function of the months (right side), Y axis is the coherence time, X axis is wind speed at 200mB on the left figure and month on the right figure and the color bar is the frequency of the coherence time.

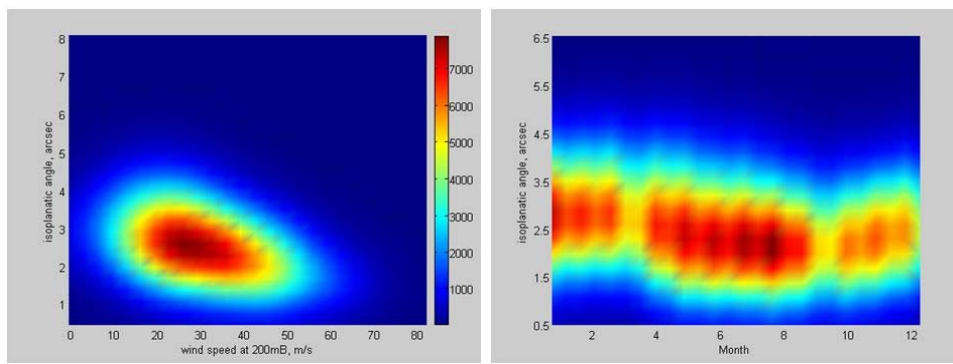

Figure 15. 2-D Histogram of the isoplanatic angle in arcsec as function of the wind speed at 200mB (left side) and the isoplanatic angle in arcsec, as function of the months (right side), Y axis is the isoplanatic angle in arcsec, X axis is wind speed at 200mB on the left figure and month on the right figure and the color bar is the frequency of the isoplanatic angle.

With this knowledge from the statistics, the Astronomers can check short and long term jet stream forecast on the web page to schedule or not AO programs for the night, according to the current conditions, in addition to the real time information provided by the Astronomical Site Monitoring facility (ASM).

AO Astronomers have developed some strategies in order to get the best performance according to the conditions when they are using the adaptive optics facilities, below an example for the Naos Conica instrument[5]

Best strategy for observations (to get the best SR):

Coherence time < 3ms is a bad value and one should not expect the best correction except at the highest frequency of the WFS and if the seeing is reasonably good.

No so good seeing(>0.6" & <1.0") & long coherence time, use a 14x14 configuration even at low frequency

Good seeing (<0.6") but short coherence time, use a 7x7 configuration at the highest possible frequency

All good use a 14x14 configuration at the highest possible frequency

All bad go to sleep

For VLTI Astronomers the Coherence Time is one or the more important parameter they have to deal followed by the Precipitable Water Vapor if they are using a mid infrared instrument like MIDI

## 6. AND FOR THE FUTURE?

Worth exploring the option of having a network of mass-dimm located a few tens of kilometers around the observatory in order to catch the turbulence traveling with the wind. As an example, Armazones is 20kms NE of Paranal (approximately) and Figure 16 shows a 20 minutes bin seeing cross correlation for 2 nights when Paranal got Armazones conditions 20 and 40min after.

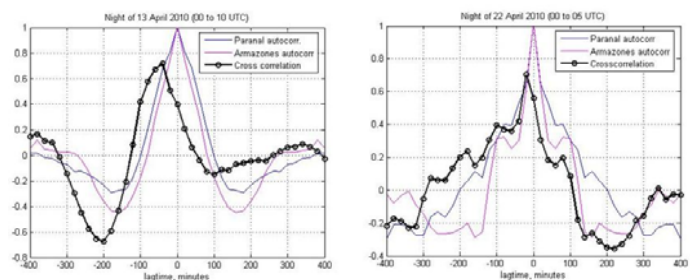

Figure 16. 20min. bin Paranal-Armazones seeing cross correlation. The time between circles is 20 minutes and in the left figure, Paranal got Armazones 40min. after while in the right figure, Paranal got Armazones 20min. after.

## 7. CONCLUSIONS

All the efforts that site testing teams put in order to characterize the atmosphere by developing and installing new and better instruments (the MASS turbulence profiler extensively uses during the ELTs site testing campaigns is a good example of this), are of the most importance for Observatories like the VLT because the information that the Science Operation people that have to deal with the atmosphere get from these instruments, allow them to improve the efficiency to manage the queue by taking in real time the best decisions on the best program to be executed according to the current conditions. Is very important to improve the predictive models mainly the ones of critical importance for adaptive optics like seeing, coherence time as the in the approaching ELT era all will be adaptive.

## REFERENCES

[1] www.eso.org/public/teles-instr/vlt.html
[2] www.pl.eso.org/paranal/director/obs-review/2009/Presentations/Day1/SciOps.pdf
[3] www.eso.org/public/teles-instr/vlt/vlt-instr.html

[4] www.eso.org/gen-fac/pubs/astclim/forecast/meteo
[5] www.eso.org/sci/facilities/paranal/sciops/team_only/NACO/Operations_Manual/NACO_OpMan.pdf